\documentclass[preprint,12pt]{elsarticle}

\usepackage{graphicx}
\usepackage{amssymb}
\journal{Physics Letters B}

\begin{document}

\begin{frontmatter}



\title{Constraints on Low-Mass WIMP Interactions on $^{19}$F from PICASSO}


\author[mtl]{S.~Archambault\fnref{mcgill}}
\author[iusb]{E.~Behnke} 
\author[saha]{P.~Bhattacharjee} 
\author[saha]{S.~Bhattacharya} 
\author[qu]{X.~Dai\fnref{chalk}}
\author[saha]{M.~Das} 
\author[qu]{A.~Davour} 
\author[mtl]{F.~Debris}
\author[lu]{N.~Dhungana}
\author[lu]{J. ~Farine} 
\author[ab]{S.~Gagnebin} 
\author[mtl]{G.~Giroux\fnref{bern}} 
\author[iusb]{E.~Grace}
\author[mtl]{C.~M.~Jackson} 
\author[qu]{A.~Kamaha}
\author[ab]{C.~Krauss} 
\author[mtl]{S.~Kumaratunga\fnref{triumf}}
\author[mtl]{M.~Lafreni\`ere}
\author[mtl]{M.~Laurin}
\author[snolab]{I.~Lawson} 
\author[mtl]{L.~Lessard} 
\author[iusb]{I.~Levine} 
\author[qu]{C.~Levy} 
\author[ab]{R.~P.~MacDonald\fnref{king}}
\author[ab]{D.~Marlisov} 
\author[mtl]{J.~-P.~Martin} 
\author[ab]{P.~Mitra} 
\author[qu]{A.~J.~Noble} 
\author[mtl]{M.~-C.~Piro\corref{cor}}
\author[lu]{R.~Podviyanuk} 
\author[prague]{S.~Pospisil} 
\author[saha]{S.~Saha} 
\author[mtl]{O.~Scallon} 
\author[saha]{S.~Seth}
\author[mtl]{N.~Starinski} 
\author[prague]{I.~Stekl} 
\author[lu]{U.~Wichoski} 
\author[qu]{T.~Xie}
\author[mtl]{V.~Zacek}
\address[mtl]{D\'epartement de Physique, Universit\'e de Montr\'eal, Montr\'eal, H3C 3J7, Canada}
\address[iusb]{Department of Physics \& Astronomy, Indiana University South Bend, South Bend, IN 46634, USA}
\address[saha]{Saha Institute of Nuclear Physics, Centre for AstroParticle Physics (CAPP), Kolkata, 700064, India}
\address[qu]{Department of Physics, Queen's University, Kingston, K7L 3N6, Canada}
\address[lu]{Department of Physics, Laurentian University, Sudbury, P3E 2C6, Canada}
\address[ab]{Department of Physics, University of Alberta, Edmonton, T6G 2G7, Canada}
\address[snolab]{SNOLAB, 1039 Regional Road 24, Lively ON, P3Y 1N2, Canada}
\address[prague]{Institute of Experimental and Applied Physics, Czech Technical University in Prague, Prague, Cz-12800, Czech Republic}
\fntext[mcgill]{Present address: Department of Physics, McGill University, Montr\'eal, H3A 2T8, Canada}
\fntext[chalk]{Present address: AECL Chalk River Laboratories, Chalk River, K0J 1J0, Canada}
\fntext[bern]{Present address: Labaratorium f\"{u}r Hochenergiephysik, Universit\"{a}t Bern, CH-3012 Bern, Switzerland}
\fntext[triumf]{Present address: TRIUMF, Vancouver, V6T 2A3, Canada}
\fntext[king]{Present address: Faculty of Science, The King's University College, Edmonton, T6B 2H3, Canada}
\cortext[cor]{Corresponding author: e-mail: piro@lps.umontreal.ca}

\begin{abstract}
Recent results from the PICASSO dark matter search experiment at SNOLAB are reported.  These results were obtained using a subset of 10 detectors with a total target mass of 0.72~kg of $^{19}$F and an exposure of 114~kgd.   The low backgrounds in PICASSO allow recoil energy thresholds as low as 1.7~keV to be obtained which results in an increased sensitivity to interactions from Weakly Interacting Massive Particles (WIMPs) with masses below 10~GeV/c$^{2}$. No dark matter signal was found. Best exclusion limits in the spin dependent sector were obtained for WIMP masses of 20~GeV/c$^{2}$ with a cross section on protons of $\sigma_{p}^{SD} = 0.032$~pb (90\% C.L.). In the spin independent sector close to the low mass region of 7~GeV/c$^{2}$ favoured by CoGeNT and DAMA/LIBRA, cross sections larger than $\sigma_{p}^{SI} = 1.41 \times 10^{-4}$~pb (90\% C.L.) are excluded.   
\end{abstract}


\end{frontmatter}

%\linenumbers

\section{Introduction}
\label{Int}
PICASSO searches for WIMP scattering using superheated liquid droplets, a variant of the bubble chamber technique~\cite{PhysRev.87.665, springerlink:10.1007/BF02781560}. The abundance of $^{19}$F in the  target liquid C$_{4}$F$_{10}$ gives PICASSO an increased sensitivity to spin dependent WIMP interactions since, with the exception of neutralino scattering on free protons,  $^{19}$F is the most favorable nucleus for direct detection. Measurements and shell model calculations of nuclear magnetic moments show the spin $1/2$ of $^{19}$F is carried almost exclusively by its unpaired proton, enhancing the spin dependent cross section by nearly an order of magnitude compared to other frequently used detector materials~\cite{Ellis1991259, PhysRevD.55.503}.  The light target nucleus $^{19}$F together with the low recoil detection threshold of $1.7$~keV render the experiment particularly sensitive to low WIMP masses below 15~GeV/c$^{2}$. This is especially interesting following the DAMA/LIBRA and recent CoGeNT and CRESST results~\cite{springerlink:10.1140/epjc/s10052-010-1303-9, PhysRevLett.106.131301,Angloher:2011uu} which are suggestive of a low mass WIMP solution of order 10~GeV/c$^{2}$. Therefore this work will explore both the implications of the new data for searches in the spin dependent sector, and the sensitivity to the low mass region in the spin independent sector. Previous results obtained with the same apparatus at SNOLAB, but using only two detectors with higher intrinsic background and with smaller exposure (14~kgd), were presented in~\cite{Archambault2009185}.

%%%%%%%%%%%%%%%%%%%%%
\section{Detector Principle}
\label{dp}
The detector medium in PICASSO is an emulsion containing C$_{4}$F$_{10}$ droplets of about 200~$\mu$m diameter in polymerized water saturated acrylamide. Since C$_{4}$F$_{10}$ has a boiling temperature of $T_{b} = -1.7$~$^{\circ}$C at a pressure of 1.013~bar, at ambient pressures and temperatures the droplets can be in a moderately metastable superheated state. A heat spike created by the energy deposition of a charged particle traversing a liquid droplet triggers a phase transition if it occurs within a certain critical length (of order tens of nm) and exceeds a certain critical energy (of order keV). Both quantities decrease exponentially with increasing temperature and are functions of surface tension, latent heat of evaporation and superheat, where the latter is defined as the difference between the vapor and external pressures of the liquid.  Details of the detector principle are explained in~\cite{RobertE1979603, Ing19971}. The phase transition is explosive and each bubble nucleation is accompanied by an acoustic signal in the audible and ultrasonic frequency range, which is recorded by piezoelectric transducers. 

\begin{figure}[htb]
\begin{center}
\includegraphics[width=.58\textwidth]{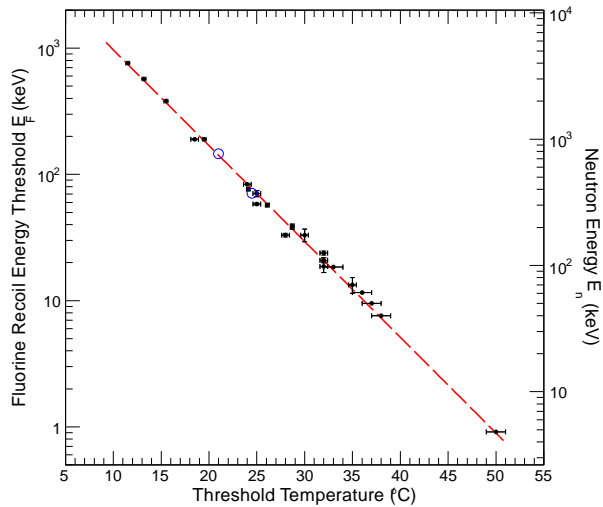}
\end{center}
\caption{Calibration curve for the energy threshold of $^{19}$F recoils as a function of temperature obtained from measurements with mono-energetic neutrons; $\alpha$ particle measurements are shown as open (blue) circles at 21~$^{\circ}$C and 25~$^{\circ}$C.}\label{Fig:1}
\end{figure}

Since the detector captures phase transitions, it performs as an energy threshold device which can be controlled by setting the temperature and/or pressure. The relation between the energy threshold E$_{th}$(T) and the operating temperature in C$_{4}$F$_{10}$ has been determined by measurements using mono-energetic neutron beams and with $\alpha$ emitters of known energies (all at 1~bar).  The results of these calibrations are shown in Fig.~\ref{Fig:1} and allow a precise description of the temperature dependence of energy thresholds ranging from 0.9~keV up to 800~keV. Details of these measurements by PICASSO can be found in~\cite{BarnabeHeider2005184, 1367-2630-13-4-043006}. The gap in the recoil energy thresholds between 0.9~keV and 7.6~keV is due to the absence of prominent resonances in the $^{51}$V(p,n)$^{51}$Cr reaction cross section used for the calibration of the low energy thresholds\footnote{Efforts are ongoing to add points across the gap by using the smaller resonances in the $^{51}$V(p,n)$^{51}$Cr cross section and especially at 5.1~keV by exploiting a resonance in the $^{19}$F-neutron cross section. In addition a point at 4.7~keV and 42~$^{\circ}$C can be inferred for C$_{4}$F$_{10}$ from neutron calibrations with C$_{4}$F$_{8}$ made by other authors~\cite{errico_rad}.}.

%%%%%%%%%%%%%%%%%%%%%
\section{Response to Different Particles}
\label{rtdps}
\begin{figure}[htb]
\begin{center}
\includegraphics[width=.8\textwidth]{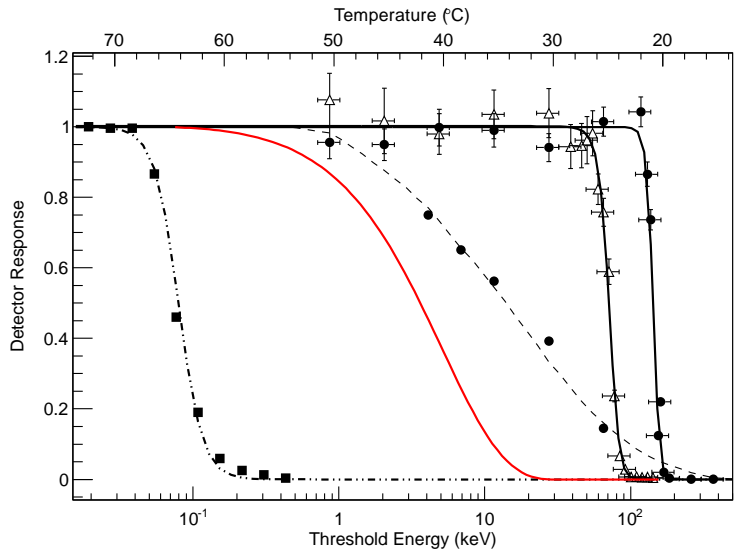}
\end{center}
\caption{Response to different kinds of particles in superheated C$_{4}$F$_{10}$. From left to right: 1.75~MeV $\gamma$-rays and minimum ionizing particles (dot-dashed); $^{19}$F recoils modeled assuming the scattering of a 50 GeV/c$^{2}$ WIMP (red); poly-energetic neutrons from an AcBe source (dotted); $\alpha$ particles at the Bragg peak from $^{241}$Am decays (open triangles); and $^{210}$Pb recoil nuclei from $^{226}$Ra spikes (full dots).}\label{Fig:2}
\end{figure}

Since each temperature at a fixed constant pressure corresponds to a defined recoil energy threshold, the spectrum of the particle induced energy depositions can be constructed by varying the temperature. A summary is shown in Fig.~\ref{Fig:2}. WIMP induced recoil energies of $^{19}$F nuclei are expected to be smaller than 100~keV and therefore become detectable above 30~$^{\circ}$C (at 1~bar). Particles which produce only low ionization densities, such as cosmic muons, $\gamma$ and $\beta$ radiation, become detectable when they create sub-keV energy clusters within sub-nm sized regions; this is only observable above 50$^{\circ}$C (less than $\approx1$~keV). These particles are well separated from strongly ionizing neutron or WIMP induced recoils, which allows efficient suppression of such backgrounds at the level of $10^{-8}$ to $10^{-10}$.  

Alpha-emitters produce a different response. In Fig.~\ref{Fig:2} the $\alpha$ curve with the lower threshold energy (higher threshold temperature) was obtained after spiking the inactive detector matrix with $^{241}$Am such that only $\alpha$ particles entering the droplets can induce nucleations. At the threshold which corresponds to a deposited energy of E$_{dep}$= 71~keV, only $\alpha$ particles with energy depositions at the Bragg peak trigger nucleation. The higher $\alpha$ energy threshold shown in Fig.~\ref{Fig:2} (full dots) was obtained with $^{226}$Ra spiked detectors. In this case the $^{226}$Ra daughter $^{222}$Rn diffuses into the droplets and the $^{210}$Pb nucleus with the highest recoil energy in the decay chain (E$_{rec} = 146$~keV) defines the threshold. As shown in Fig.~\ref{Fig:1}, the $^{226}$Ra and $^{241}$Am related thresholds (open circles) are found to be in good agreement with the energy thresholds obtained in the calibrations with mono-energetic neutrons. In both cases, if the energy thresholds decrease below E$_{dep} = 71$~keV the liquid becomes sensitive to smaller dE/dx on the $\alpha$ track. It is important to note in Fig.~\ref{Fig:2}, that the response remains flat from $1-120$~keV. This has been confirmed with numerous detectors with large $\alpha$ background and indicates that the detectors are within an uncertainty of less than 3\% fully sensitive to energy depositions above threshold. A more detailed discussion can be found in~\cite{1367-2630-13-4-043006}. 

Since the detectors are fully sensitive to $\alpha$ particles over the entire range of the WIMP sensitivity, $\alpha$ particles are the most important background for this kind of dark matter search. However the shapes of the WIMP (essentially exponentially falling) and of the $\alpha$ (constant) responses differ substantially, such that they can be separated by fitting the two contributions (Sect.~\ref{dcaa}).

%%%%%%%%%%%%%%%%%%%%%
\section{Experimental Setup}
\label{es}
The present PICASSO installation at SNOLAB accommodates 32 detector modules. The detectors are installed in groups of four inside thermally and acoustically insulated chambers, serving as a temperature control unit with a precision of  $\pm 0.1$~$^{\circ}$C in the range from 20~$^{\circ}$C to 50~$^{\circ}$C. The current detector generation consist of cylindrical modules of 14~cm diameter and 40~cm height~\cite{Archambault2009185}. The containers are fabricated from acrylic and are closed on top by stainless steel lids sealed with polyurethane O-rings. Each detector is filled with 4.5 litres of polymerized emulsion loaded with droplets of C$_{4}$F$_{10}$. The active part of each detector is topped by mineral oil, which is connected to a hydraulic manifold in order to allow periodic pressurizations of the detectors to reconvert bubbles back into droplets.

 In the most recent detector generation, the emulsion has glycerine and polyethylene glycol as the main ingredients. During fabrication the viscosity of the non-polymerized liquid is used to suspend the C$_{4}$F$_{10}$ droplets homogeneously and uniformly. The volume distribution of droplets peaks at diameters of around 200~$\mu$m. On average the active mass of a detector used in this analysis is 90~g of C$_{4}$F$_{10}$ corresponding to 72~g of $^{19}$F. The active mass is known with a precision of 1\% from weighing during fabrication, but additional uncertainties might arise due to losses of C$_{4}$F$_{10}$ during polymerization or by diffusion into the matrix. Therefore the active detector masses and sensitivities are verified and monitored by measurements with a calibrated AmBe neutron source. The values quoted in Table~\ref{Tab:1} for the detectors used in this analysis are the averages of the $^{19}$F masses determined during fabrication and neutron calibration measurements. The quoted errors are: individual detector uncertainties from weighing and calibration during and after the fabrication process; and a common systematic uncertainty from calibrations with the poly-energetic neutron sources in the lab and underground (AcBe/AmBe).  A description of the fabrication and purification of this type of detector can be found in~\cite{Piro-ICATPP2010}. 
 \begin{table}
\begin{center}
\begin{tabular}{| c | c c  |}
\hline
Detector & Mass & Exposure  \\
 & $g(F)$ & $kg(F)d$  \\
 \hline
 71 & $64.66 \pm 2.40 \pm 1.94$ & $16.09 \pm 0.77$ \\
 72 & $59.87 \pm 1.60 \pm 1.80$ & $17.69 \pm 0.71$ \\
 131 & $82.79 \pm 3.11 \pm 2.8$ & $10.89 \pm 0.55$ \\
 134 & $71.61 \pm 0.80 \pm 2.15$ & $15.94 \pm 0.51$ \\
 137 & $81.35 \pm 2.56 \pm 2.44$ & $16.33 \pm 0.71$ \\
 141 & $68.70 \pm 2.88 \pm 2.06$ & $13.37 \pm 0.69$ \\
 144 & $41.51 \pm 1.60 \pm 1.42$ & $6.18 \pm 0.31$ \\
 145 & $69.85 \pm 2.79 \pm 2.10$ & $7.83 \pm 0.39$ \\
 147 & $66.26 \pm 2.63 \pm 1.99$ & $6.55 \pm 0.32$ \\
 148 & $109.53 \pm 3.27 \pm 3.3$ & $3.43 \pm 0.15$ \\
 \hline
 \end{tabular}
 \caption{Summary of the performance parameters of all detectors used in this analysis. Active masses are normalized to the mass of $^{19}$F present in a module. The quoted mass errors are: individual uncertainties from weighing and neutron calibration during and after the fabrication process; and a common systematic uncertainty from calibrations with a  poly-energetic neutron source (AmBe). The indicated values for exposure cover data taken over the entire temperature range from 28~$^{\circ}$C to 48~$^{\circ}$C. }\label{Tab:1}
 \end{center}
 \end{table}

Each detector module has nine piezo-electric transducers, mounted at three different heights on a flat spot, milled into the outside of the acrylic container wall. The transducers are ceramic disks (Ferroperm PZ27) with a diameter of 16~mm and 8.7~mm thickness and a pressure sensitivity of 27~$\mu$V/$\mu$bar.  The piezoelectric sensors are read out by custom made low-noise preamplifiers. Details of the electronic read-out are reported in~\cite{martin-picassodaq}.  The trigger has multiplicity one: triggering of any of the nine channels causes all channels to acquire data. The trigger is fully sensitive at temperatures above 24~$^{\circ}$C (recoil energies smaller than 78~keV). One detector in the set up, not loaded with active liquid but fabricated and read out as the others, serves as a monitor for non-particle related backgrounds such as mine blasts, electronic spikes, detector cross-talks and ambient noise sources. 

The entire installation is surrounded by a 30.5~cm thick water shield, which serves as a neutron moderator and absorber. This shielding is made of 242 cardboard boxes containing square polyethylene bags filled with water, with a filling factor of about 75\%. At the location of the experiment, a depth of 2070~m, 90\% of the fast neutrons above 5~keV are produced by ($\alpha$, n) reactions in the surrounding Norite rock, with the remaining 10\% being fission neutrons. The fast neutron flux was measured to be $\sim 3000$ neutrons m$^{-2}$d$^{-1}$~\cite{SNO-background}. In order to estimate the expected neutron flux reduction by the shielding, Monte-Carlo (MC) simulations have been performed which included all structural materials, the geometric filling factor of the water boxes and self shielding effects due to the presence of other detectors within the shielding. The performance of the MC simulation was checked against measurements using $^{3}$He counters (SNO NCD-counters) and good agreement was found. The simulations predict a reduction of fast neutrons from the shielding by a factor of 35, giving an estimated neutron induced count rate at the level of 1.1 neutrons kg$^{-1}$d$^{-1}$ (kg of $^{19}$F) for operation at 5~keV threshold energy. %Possible uncertainties in this estimate come from the filling factor and environmental effects close to the set up, such as the presence of a large water storage tank, both of which might result in a further reduction of the predicted neutron flux.

%%%%%%%%%%%%%%%%%%%%%
\section{Acoustic Signatures for Background Reduction}
\label{asfbr}
Apart from the different temperature or threshold energy profiles which can be used to discriminate different particle interactions in superheated liquids (Fig.~\ref{Fig:2}), the acoustic signals themselves can be exploited for the discrimination of particle and non-particle sources. Calibrations with neutron test beams and fast neutrons from AcBe/AmBe sources show that the associated waveforms have characteristic frequency and time dependences: a short rise time, reaching a maximum after 20-40~$\mu$s, with slower oscillations following for several milliseconds. In addition the amplitude distributions of the high frequency content ($ > 18$~kHz) of the particle induced wave forms are concentrated in a well defined peak. These features are used to construct variables which allow the discrimination of particle induced events from non-particle backgrounds. 

\emph{Acoustic energy (EVAR)}:  This parameter measures the acoustic energy of an event. Frequencies below 18 kHz were found to carry no relevant  information and are removed by a Butterworth high pass filter applied to the Fourier transformed acoustic signal. The waveform is squared and integrated over the signal duration, starting from a fixed pre-trigger time. The resulting values are then averaged over all active transducers to reduce solid angle effects. The resolution at FWHM is $\sim 20$ \% for temperatures tested, while the centre of the distribution increases smoothly with temperature. The parameter \emph{EVAR} is used to define an acoustic energy threshold to stay sufficiently away from non-particle related noise signals. Details concerning the underlying physics processes are described in~\cite{1367-2630-13-4-043006, 1367-2630-10-10-103017}. Since the expected signals from WIMP induced recoils have a similar intensity to neutron recoils this discrimination variable is of prime importance for dark matter searches with superheated liquids. 

\emph{Frequency content (FVAR)}: Studies of the fast Fourier transforms (FFT) of particle induced waveforms have shown that the majority of the signal power can be found in the frequency range between 20 and 70~kHz. A variable \emph{FVAR} is constructed by taking the logarithm of the ratio of signal power in the intervals from 20-30~kHz and 45-55~kHz. This variable allows suppression of fractures or secondary events which have a significant deficit in signal power in the low frequency window; these events sometimes follow a true particle induced event and are caused by a weakening of the matrix. Mine blast events are also  efficiently removed by cuts applied on this variable. A more detailed discussion of the discrimination variables and the event types they are able to discriminate can be found in~\cite{Archambault2009185} and especially Fig.~4 therein.

\begin{figure}[htb]
\begin{center}
\includegraphics[width=.8\textwidth]{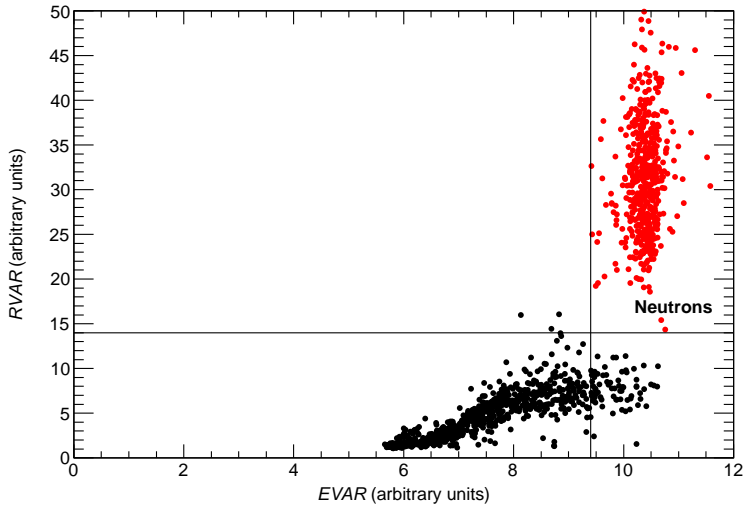}
\end{center}
\caption{The signal energy (\emph{EVAR}) and rise time (\emph{RVAR}) related variables allow the discrimination of particle induced events from other acoustic signals created by activities in the detector matrix. Data taken at 45~$^{\circ}$C  during calibration runs are shown. Neutron induced events cluster in the upper right rectangle with well defined \emph{EVAR} and background events concentrate at low values of \emph{RVAR}.}\label{Fig:3}
\end{figure}

\emph{Signal rise time (RVAR)}:  This parameter reflects the steepness of growth of the signals and measures the energy content within the first 25~$\mu$s after the signal start time. This variable was introduced in order to suppress a class of background events with a characteristic slow rise time, but with an acoustic energy and frequency content comparable to particle induced events. This background became noticeable in detectors with increasingly smaller intrinsic $\alpha$ contamination, especially above 40~$^{\circ}$C where this background increases nearly exponentially. The most probable cause of these events is a cascade of secondary vaporizations in the vicinity of primary particle induced events.  Since these signals have only slightly reduced contribution at lower frequencies, they can only be partially removed by the \emph{FVAR} variable.  A scatter plot of the variables \emph{RVAR} vs.\ \emph{EVAR} at 45~$^{\circ}$C is shown in Fig.~\ref{Fig:3}. Particle induced events accumulate in the right upper rectangle, secondary background events concentrate at low values of \emph{RVAR}.

%%%%%%%%%%%%%%%%%%%%%
\section{Data Collection and Analysis}
\label{dcaa}
\begin{figure}[htb]
\begin{center}
\includegraphics[width=.8\textwidth]{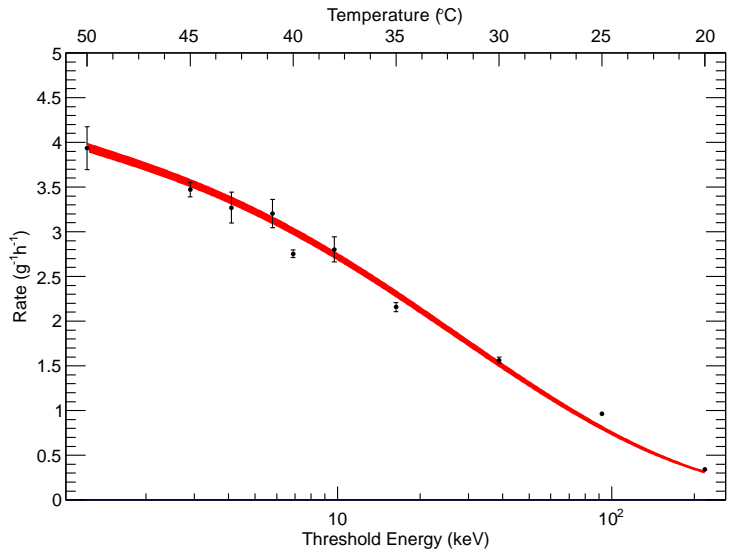}
\end{center}
\caption{Combined data from all detectors from calibration runs with poly-energetic neutrons (AmBe). Data were taken in regular intervals spread over the entire data taking period. For a given temperature, data from all detectors have been corrected for cut efficiencies, combined in a weighted average and are compared to simulations (red). The threshold energy scale refers to $^{19}$F recoils; for recoiling $^{12}$C nuclei, energies have to be multiplied by 1.47. Uncertainties shown are statistical only; the observed scatter of some points is due to the uncertainty in the location of the neutron source which introduces a systematic uncertainty at the level of 5\% at each temperature point}\label{Fig:4}
\end{figure}

The analysis presented here was performed on a group of 10 detectors. Seven of these detectors were installed at the end of 2008 and belong to the most recent generation of PICASSO detectors with the lowest internal background and with sufficient exposure to contribute significantly to the analysis. The remaining three detectors belong to the previous generation and were continuously taking data from June 2007; of these, two are the detectors used in the 2009 analysis~\cite{Archambault2009185}.  A WIMP run typically lasts 40 hours after which the detectors are recompressed for 15 hours at a pressure of 6 bar in order to reduce bubbles to droplets and to prevent excessive bubble growth which could damage the polymer. A total of 264 WIMP runs were analysed within this period yielding a total exposure of 114.3~kgd in the background and signal regions. Approximately every three months calibration data have been taken at several temperatures with a weak AmBe neutron source ($68.71 \pm 0.74$~s$^{-1}$), placed equidistant at $10 \pm 2$~cm from the centre of each detector~\cite{loach}.  These data were used to monitor the stability of the detectors and to determine cut efficiencies for the discrimination variables \emph{EVAR}, \emph{FVAR} and \emph{RVAR}. The combined data from all detector calibration runs covering the analysis period are shown in Fig.~\ref{Fig:4}. For a given temperature, data from all detectors have been combined in a weighted average and compared to MC simulations (red curve in Fig.~\ref{Fig:4}). The observed scatter in some of the data points is caused by the uncertainty of the source position which introduces an additional systematic uncertainty at the level of 5\% at each temperature point. These tests monitor the long term stability of the detectors. They demonstrate that once the count rates have been normalized by grams of C$_{4}$F$_{10}$ the entire detector array behaves consistently, as one large detector.

The analysis proceeds in the following order: 
\begin{itemize}
\item A list of golden runs is established for each detector. To qualify as a golden run: at least six working acoustic readout channels are required; the duration of the run must exceed 15~h (1~h for calibrations); and the gauge pressure in the detector has to be within 0.1~bar of the ambient pressure.  
\item A selection to remove event bursts with $< 3$ seconds ($< 0.1$~seconds for calibrations) between successive triggers is applied. In these low background detectors, the probability for successive events within 3 seconds is negligibly small, and these events are typically retriggers of the same events or events physically induced in the detector by primary expansion. 
\item An event selection is performed on \emph{EVAR}. This selection is fixed for each temperature by fitting a Gaussian curve to calibration data and by interpolating between the calibration temperatures. As a large quantity of bubbles in the matrix leads to decreasing signal amplitude, for calibration runs only the first 200 neutron induced events are selected in order to maintain acoustic conditions. Selection values are set to give 95\% acceptance. 
\item The events have to pass a selection on \emph{RVAR}, chosen to yield 95\% acceptance on calibration data. 
\item Finally the events have to pass a selection on \emph{FVAR}, again to yield 95\% acceptance on calibration data. 
\end{itemize}
The effects of the applied cuts for two temperatures on the trigger rates are illustrated for one of the detectors (144) in Table~\ref{Tab:2}. The cut on the acoustic energy variable is the most effective discriminator of non-particle related signals. The background increases with increasing temperature and the cut on the rise-time variable \emph{RVAR} becomes more important at higher temperatures.
\begin{table}
\begin{center}
\begin{tabular}{ | c | c c  | }
 \hline
  Detector  144 & 30~$^{\circ}$C & 45~$^{\circ}$C \\
 \hline
  Triggers/day & $23.4 \pm 0.9$ & $60.5 \pm 1.2$ \\
  After 3 sec cut & $15.8 \pm 0.8$ & $40.0 \pm 1.0$ \\
  After \emph{EVAR} cut & $2.3 \pm 0.3$ & $3.2 \pm 0.3$ \\
  After \emph{RVAR} cut & $2.2 \pm 0.3$ & $2.4 \pm 0.3$ \\
  After \emph{FVAR} cut & $2.1 \pm 0.3$ & $2.2 \pm 0.2$ \\
 \hline
\end{tabular}
\caption{Effect of the applied cuts on the trigger rate at 30~$^{\circ}$C and 45~$^{\circ}$C. Detector 144 is shown as an example. }\label{Tab:2}
\end{center}
\end{table}

After correcting for cut acceptances and dead time, the events recorded by the detectors at each temperature are normalized with respect to the active mass ($^{19}$F) and data taking time. The count rates of all detectors are flat in the range from 1.7 to 92~keV (48~$^{\circ}$C to 25~$^{\circ}$C), similar to that observed in the presence of $\alpha$ emitters in the droplets (Fig.~\ref{Fig:2}). The count rates averaged over this plateau range are given in Table~\ref{Tab:3} and are indicative of the level of $\alpha$ contamination in the individual detectors. The decreasing rate as a function of detector number reflects the progress in purification during fabrication over time.
\begin{table}
\begin{center}
\begin{tabular}{| c | c c c |}
\hline
Detector & Rate & $\sigma_{F}^{min}$ & $M_{W}^{min}$ \\
 & $cts/kg(F)/d$ & $pb$ & $GeV/c^{2}$  \\
 \hline
 71 & $327.6 \pm 4.3 \pm 21.6$ & $-15.43 \pm 8.71 \pm 1.4$ & 10 \\
 72  & $134.2 \pm 2.9 \pm 8.8$ & $+10.48 \pm 7.82 \pm 1.0$  & 9 \\
 131 & $31.5 \pm 1.6 \pm 2.3$ & $-1.80 \pm 3.38 \pm 0.31$  & 9 \\
 134 & $209.6 \pm 3.9 \pm 12.8$ & $+4.65 \pm 9.49 \pm 0.76$ & 7 \\
 137 & $69.9 \pm 2.1 \pm 4.7$ & $+2.76 \pm 5.44 \pm 0.48$ & 10 \\
 141 & $25.2 \pm 1.4 \pm 1.8$ & $-4.71 \pm 3.53 \pm 0.19$ & 12 \\
 144 & $60.8 \pm 3.3 \pm 4.3$ & $+1.69 \pm 6.48 \pm 0.54$ & 9 \\
 145 & $31.5 \pm 2.1 \pm 12.3$ & $-0.78 \pm 5.24 \pm 0.42$ & 12 \\
 147 & $20.6 \pm 1.8 \pm 1.5$ & $-0.86 \pm 3.01 \pm 0.26$ & 10 \\
 148 & $20.0 \pm 1.9 \pm 1.3$ & $-0.28 \pm 4.30 \pm 0.33$ & 8 \\
 \hline
 \end{tabular}
 \caption{Summary of analysis results. The averaged rates are corrected for cut efficiencies and the systematic errors reflect uncertainties in the mass determination, the detection efficiency and the cut-efficiency errors. Cross section values for WIMP interactions on $^{19}$F are quoted for a resolution parameter $a=5$ (Sect.~\ref{sfadms}) at maximum sensitivity of the fits obtained for the WIMP mass given in the corresponding column to the right. The sources of systematic uncertainties correspond to those listed in Sect.~\ref{sfadms}.}\label{Tab:3}
 \end{center}
 \end{table}  

The origin of the $\alpha$ background is still uncertain and under investigation. It seems probable that $\alpha$ emission occurs within the droplets. This hypothesis is supported by studies of the acoustic energy parameter, which show for most detectors at 30~$^{\circ}$C indications of two groups of events: one characteristic for single nucleation by the recoiling $\alpha$ emitter; and a second group of events where the recoiling nucleus and $\alpha$ particle add their contributions to the acoustic signal. As discussed in~\cite{1367-2630-13-4-043006}, at 30~$^{\circ}$C this feature is typical for detectors where the $\alpha$ emitters are located inside the droplets. Possible scenarios are either a direct contamination of the C$_{4}$F$_{10}$ itself or diffusion of $^{222}$Rn from $^{226}$Ra in the polymer matrix into the droplets. Taking as an example detector 148, with the lowest background rate, a contamination at the level of $2 \times 10^{-11}$~gU~g$^{-1}$ if the activity is located in the C$_{4}$F$_{10}$ and of $2 \times 10^{-12}$~gU~g$^{-1}$ for a contamination originating in the matrix is expected.

\begin{figure}[htb]
\begin{center}
\includegraphics[width=.8\textwidth]{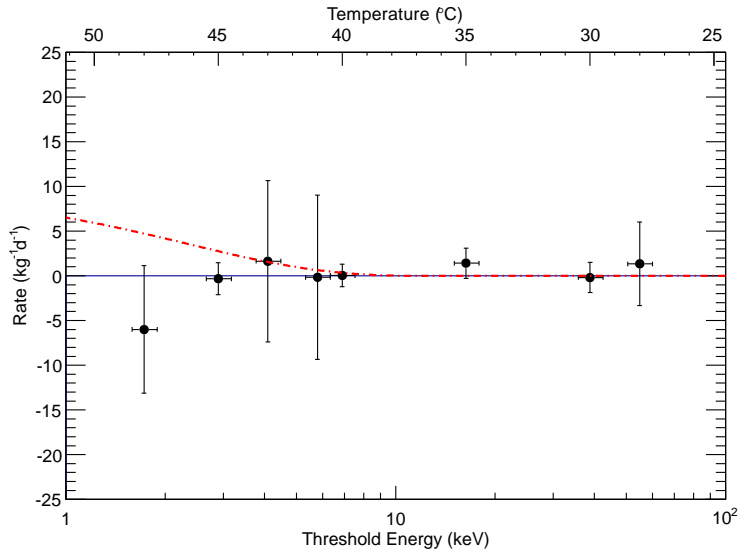}
\end{center}
\caption{Combined data from all detectors for WIMP runs. For each detector the average count rate is calculated over the entire temperature/energy range (28~$^{\circ}$C~$< T <$~48~$^{\circ}$C) and subtracted from the individual data points. Data for each detector are then combined at each temperature in a weighted average. The rate expected for a hypothetical WIMP with M$_{W} = 7$~GeV/c$^{2}$ and $\sigma^{SI}_{p} = 1.2 \times 10^{-4}$~pb is shown by the red-dotted curve.}\label{Fig:5}
\end{figure}

In order to combine for illustrative purposes all detectors in a single plot of rate vs.\ threshold energy, the data of individual detectors are renormalized by their respective $\alpha$ contamination, so that the data can be combined. For this the following procedure is carried out: for each detector the average count rate over the entire plateau temperature range is calculated (28~$^{\circ}$C~$< T <$~48~$^{\circ}$C); under the hypothesis of absence of WIMPs, this count rate is taken as an approximation of the $\alpha$ background level of the detector and is subtracted from individual data points at different temperatures; data for each detector and temperature are then combined in a weighted average; finally temperatures are converted into threshold energies, by taking into account that due to the elevated mine pressure (1.2~bar) the measured temperature at the location of the experiment corresponds to temperatures at surface where the threshold values were calibrated, reduced by 2~$^{\circ}$C. The resulting threshold energy spectrum shown in Fig.~\ref{Fig:5} exhibits several noteworthy features: the count rates of all detectors as a function of energy are essentially constant; the sensitivity of the experiment for WIMP induced deviations from the constant background is at the level of a few cts~kg$^{-1}$~d$^{-1}$ (kg $^{19}$F); for modest changes in temperature from 28~$^{\circ}$C~$< T <$~48~$^{\circ}$C the dynamic range in threshold energy sensitivity is large and covers the region from $1.7-55$~keV; errors are dominated by statistics and reflect the time spent at respective temperatures;  and in terms of sensitivity to light mass WIMPs the experiment could still gain substantially by running at the highest temperatures. Although the background is subtracted here to better visualize the spectrum, a flat background component is included in the overall fit to the spectrum during the WIMP analysis.

%%%%%%%%%%%%%%%%%%%%%
\section{Search for a Dark Matter Signal}
\label{sfadms}
To search for a dark matter signal the measured rates as a function of threshold energy are compared to those predicted for interactions of WIMPs in our galactic halo with $^{19}$F nuclei, in the presence of a constant $\alpha$ background in the detector.  We use the formalism described in~\cite{Lewin199687} which approximates the recoil energy spectrum as an exponentially falling distribution: 
\begin{equation}\label{eq:2}
\frac{dR}{dE_{R}} \approx c_{1} \frac{ R_{0} }{ \langle E_{R} \rangle } F^{2}(E_{R}) \exp\left( - \frac{ c_{2} E_{R} }{ \langle E_{R} \rangle } \right) (\mathrm{keV^{-1}kg^{-1}d^{-1}}),
\end{equation}
where $\langle E_{R} \rangle = 2 M_{F}M^{2}_{W}/(M_{F}+M_{W})^{2} \langle v^{2}_{W} \rangle$  is the mean average recoil energy; M$_{F}$ and M$_{W}$ are the masses of the $^{19}$F nucleus and of the WIMP, respectively;  $\langle v_{W} \rangle$  is the average velocity of halo dark matter particles and F$^{2}$(E$_{R})$ is a nuclear form factor taken as 1 for a light nucleus such as fluorine with small momentum transfer~\cite{Lewin199687}; and the constants $c_{1,2}$ describe the effect of the Earth's velocity,  $v_{e}$, relative to the halo ($c_{1} = 0.75$, $c_{2} = 0.56$ for $v_{e} = 244$~km s$^{-1}$). R$_{0}$ is the expected total WIMP interaction rate per kg of $^{19}$F per day,
\begin{equation}\label{eq:3}
R_{0}(M_{W},\sigma_{F})=\frac{405}{A_{T}M_{W}} \left(\frac{\sigma_{F}}{\mathrm{pb}}\right) \left(\frac{\rho_{W}}{0.3\: \mathrm{GeV\, cm^{-3}}}\right) \left(\frac{\langle v_{W} \rangle}{230\: \mathrm{km\, s^{-1}}}\right)  (\mathrm{kg^{-1} d^{-1}}),
\end{equation}
where $A_{T} =19$ is the atomic mass of the target atom; $\rho_{W}$ is the mass density of WIMPs; and $\sigma_{F}$ is the WIMP interaction cross section on $^{19}$F. Since the detector operates as a threshold device the observed rate at a given recoil energy threshold $E_{R_{th}}(T)$ is given by:
\begin{equation}\label{eq:4}
R_{obs}(M_{W},\sigma_{F},E_{R_{th}}(T)) = \int_{0}^{E_{R_{max}}} P(E_{R}, E_{R_{th}}(T)) \frac{dR}{dE_{R}} dE_{R},
\end{equation}
where $P(E_{R}, E_{R_{th}}(T))$ describes the effect of a finite resolution at threshold and the integral extends to $E_{R_{max}}$, the maximum recoil energy a WIMP can transfer at its galactic escape velocity of $v_{esc}= 600$~km s$^{-1}$. The shape of the threshold curve is discussed in more detail in~\cite{BarnabeHeider2005184,1367-2630-13-4-043006}. It is determined by calibrations with neutron sources and $\alpha$ emitters (Fig.~\ref{Fig:2}) and can be well approximated by:
\begin{equation}\label{eq:5}
P(E_{R}, E_{R_{th}}(T)) = 1 - \exp\left(a(T) \left(1- \frac{E_{R}}{E_{R_{th}}(T)}\right)\right) . 
\end{equation}
The parameter $a(T)$ describes the steepness of the energy threshold. It is related to the intrinsic energy resolution and reflects the statistical nature of the energy deposition and its conversion into heat. The larger is $a$, the steeper is the threshold. Measurements with $\alpha$ emitters gave $a = 10 \pm 1$ at 146~keV. Alpha particles depositing their energy at the Bragg peak seem to produce a less steep threshold with $a = 5.8 \pm 0.7$ at 71~keV. Measurements by other authors with 17~keV mono-energetic recoils following the reaction $^{35}$Cl(n$_{th}$,p)$^{35}$S are compatible with $1 < a < 5$~\cite{errico_rad}, but with large uncertainties. In this analysis we adopt a principal value of $a = 5$ and let the parameter vary within the interval $2.5 < a < 7.5$.

Since for $M_{W}$ smaller than 500~GeV/c$^{2}$ the response curves differ in shape from the flat $\alpha$ background of each detector, an upper bound on $\sigma_{F}$ is obtained for each individual detector by fitting the WIMP response curve and the flat $\alpha$ background. For a given $M_{W}$ the two parameters of the fit are $\sigma_{F}$ and a scale factor describing the constant background. The result for each detector is shown in Table~\ref{Tab:3}. Combined in a weighted average, the maximum sensitivity occurs for WIMPs in the mass region around $M_{W} = 10$~GeV/c$^{2}$ and with $\sigma_{F}  = -0.72 \pm 1.45 \pm 0.12$~pb (1 standard deviation; $a=5$); this null result can be converted into a limit~\cite{PhysRevD.57.3873} on the cross section for WIMP interactions on $^{19}$F of $\sigma_{F} = 2.00$~pb (90\% C.L.) for resolution parameter $a=5$. 

The main systematic uncertainties (1 standard deviation) affecting these limits on $\sigma_{F}$ are in order of importance:
\begin{itemize}
\item a 3\% common systematic uncertainty in the determination of the active mass of the detectors, resulting in a 3\% uncertainty in the cross-section limit; 
\item a 3\% uncertainty in the recoil detection efficiency inferred from the response of $\alpha$ particles; 
\item a 2.5\% uncertainty in the \emph{EVAR} cut acceptance and a 1.5\% uncertainty due to curve fitting of \emph{EVAR}, results in a 3\% uncertainty in the limit;
\item similarly the event selection results in a 3\% uncertainty from \emph{RVAR} and a 3\% uncertainty from \emph{FVAR};  
\item a 1~$^{\circ}$C systematic shift in temperature during test beam calibrations would result in an energy scale shift, introducing a 1\% uncertainty in the cross section limits; 
\item atmospheric pressure changes at the level of 3\% result in uncertainties $<1$\%;  
\item and the hydrostatic pressure gradient of $\pm 2$\% with respect to the centre of a detector module can be translated into an uncertainty of $<1$\% in the cross section.      
\end{itemize}
The variation of the energy resolution parameter within the uncertainty range $a = 5 \pm 2.5$ results in a $\pm 1.5$\% change in the cross section limit at 10~GeV/c$^{2}$. This uncertainty increases at lower WIMP masses and is shown as a broadening of the limits into confidence bands (Sects.~\ref{litsds} and~\ref{litsis}).

%%%%%%%%%%%%%%%%%%%%%
\section{Limits in the Spin Dependent Sector}
\label{litsds}
The interaction of dark matter particles with nuclei of ordinary matter of electro-weak strength has the general form: 
\begin{equation}\label{eq:6} 
\sigma_{A} = 4 G^{2}_{F} \left(\frac{M_{W}M_{A}}{M_{W}+M_{A}}\right)^{2} C_{A}F(q^{2}),
\end{equation}         
where $G_{F}$ is the Fermi constant, and $M_{W,A}$ are the masses of the WIMP and detector nuclei respectively~\cite{Jungman1996195}. $C_{A}$ is an enhancement factor dependent on the type of WIMP interaction and $F(q^{2})$ is a nuclear form factor which becomes important for large mass number, $A$, and momentum transfer, $q$. 

\begin{figure}[htb]
\begin{center}
\includegraphics[width=.8\textwidth]{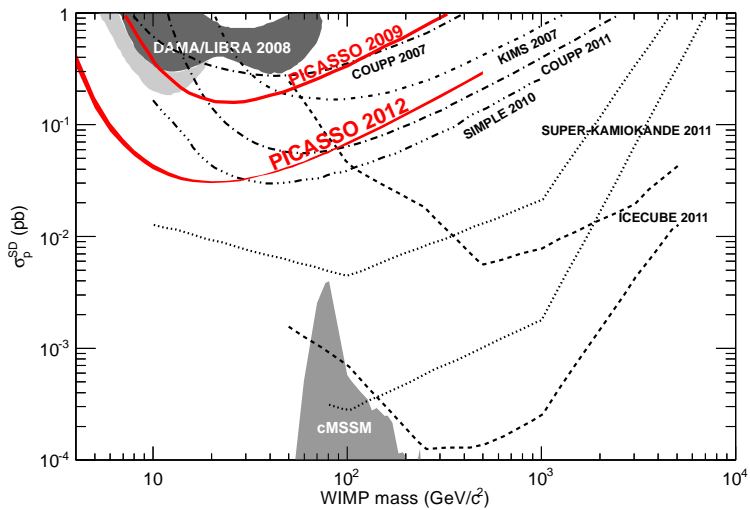}
\end{center}
\caption{Upper limits at 90\% C.L. on spin dependent WIMP-proton interactions. PICASSO limits are shown as full lines. Additional curves are from KIMS~\cite{PhysRevLett.99.091301}, COUPP~\cite{PhysRevLett.106.021303} and SIMPLE~\cite{PhysRevLett.105.211301}\protect\footnotemark. The DAMA/LIBRA~\cite{springerlink:10.1140/epjc/s10052-010-1303-9,1475-7516-2009-04-010} allowed regions are also shown (light grey: with ion channelling). Also shown are the spin dependent search results in both soft and hard annihilation channels from SuperK~\cite{0004-637X-742-2-78} and AMANDA-II/IceCube~\cite{IceCube-2011}; and theoretical predictions discussed in~\cite{1126-6708-2007-07-075,PhysRevD.63.065016}.}\label{Fig:6}
\end{figure}

Spin dependent interactions (SD) with axial vector couplings involve squark and Z exchanges and depend on the spin of the target nucleus with an enhancement factor of the form: 
\begin{equation}\label{eq:7}
C^{SD}_{A} =  \frac{8}{\pi} \left[a_{p} \langle S_{p} \rangle + a_{n} \langle S_{n} \rangle \right]^{2} \frac{J+1}{J},
\end{equation}
 where $a_{p,n}$ are the effective proton (neutron) coupling strengths, $\langle S_{p,n} \rangle$ are the expectation values for the nucleon spins in the target nucleus ($\langle S_{p} \rangle = 0.44$ and $\langle S_{n} \rangle = -0.19$ in $^{19}F$) and $J$ is the nuclear spin~\cite{Jungman1996195, Engel-1-1992, PhysRevD.40.2131}.  Assuming that scattering of dark matter on $^{19}$F is dominated by interactions with protons, the cross section $\sigma_{p}^{SD}$ for scattering on protons is related to $\sigma_{F}$ by:
\begin{equation}\label{eq:8}
\sigma_{p}^{SD} = \sigma_{F} \left(\frac{\mu_{p}}{\mu_{F}}\right)^2 \frac{C_{p}^{SD}}{C_{p(F)}^{SD}}.
\end{equation}
Here $\mu_{p,F}$  are the WIMP-proton (fluorine) reduced masses, $C_{p}^{SD}$ is the enhancement factor for scattering on the free proton and $C_{p(F)}^{SD}$ is the corresponding quantity for scattering on protons in the $^{19}$F nucleus. $C_{p(F)}^{SD}$ is obtained by setting $a_{n} = 0$ in Eq.~\ref{eq:7} and yields the ratio $C_{p}^{SD}/C_{p(F)}^{SD}  = 1.285$~\cite{Tovey200017, PhysRevLett.93.161301}. With Eq.~\ref{eq:8} the fit result for $\sigma_{F}$ can be converted into a cross section on protons of $\sigma_{p}^{SD} = -0.008 \pm 0.022 \pm 0.002$~pb (1 standard deviation; $a=5$), yielding a best limit of $\sigma_{p}^{SD} = 0.032$~pb (90\% C.L.) for WIMP masses around 20~GeV/c$^{2}$.  The resulting exclusion curve for the WIMP cross section on protons as a function of WIMP mass is shown in Fig.~\ref{Fig:6} together with published results in the spin dependent sector. The broadening of the exclusion curve shows the effect of varying the energy resolution parameter $a$ within its uncertainty.
\addtocounter{footnote}{-1}
\footnotetext{The SIMPLE collaboration has recently claimed very competitive limits in arXiv:1106.3014; see, however, arXiv:1106.3559 and arXiv:1107.1515. }

%%%%%%%%%%%%%%%%%%%%%
\section{Limits in the Spin Independent Sector}
\label{litsis}
Spin independent (SI) or scalar interactions proceed via Higgs  and/or squark exchanges, with $C_{A}$ in Eq.~\ref{eq:6} given by:
\begin{equation}\label{eq:9}
C_{A}^{SI} = \frac{1}{4 \pi} \left[Z f_{p} + (A - Z) f_{n}\right]^{2},
\end{equation}
where $f_{n,p}$ are the WIMP couplings to the nucleons. For equal couplings to neutrons and protons the cross section is proportional to $A^{2}$ (coherent interaction). Using this assumption the cross section becomes:  
\begin{equation}\label{eq:10}
\sigma_{p}^{SI} = \sigma_{F} \left(\frac{\mu_{p}}{\mu_{F}}\right)^{2} \frac{1}{A^{2}},
\end{equation}
with $A= 19$. The limits on $\sigma_{F}$  can be translated into an upper bound on the WIMP proton cross section in the spin independent sector, with maximum sensitivity at $M_{W} = 20$~ GeV/c$^{2}$ and  $\sigma_{p}^{SI}  = 6.1 \times 10^{-5}$~pb (90\% C.L.; $a=5$). The effect of scattering on $^{12}$C nuclei in the target, including a shift in the energy threshold, is estimated to be of order 10\% and is included in the results.   

At the maximum sensitivity these limits are three orders of magnitude less stringent than the best limits reached by XENON100 and CDMS in the SI sector~\cite{PhysRevLett.107.131302,PhysRevLett.106.131302} in the range of 50~GeV/c$^{2} < M_{W} < 80$~GeV/c$^{2}$. However for low mass dark matter particles ($M_{W} < 10$~GeV/c$^{2}$) and heavy target nuclei the advantage of coherent scattering in SI interactions is largely lost and comparable sensitivity can be obtained with a light mass target nucleus, such as $^{19}$F, combined with a low energy detection threshold. This low mass region has become especially interesting  in view of the DAMA/LIBRA and recent CoGeNT results which indicate an annual modulation effect for a WIMP with a mass of 7~GeV/c$^{2}$ and a SI cross section close to $1.2 \times 10^{-4}$~pb~\cite{springerlink:10.1140/epjc/s10052-010-1303-9, PhysRevLett.106.131301}. In the same mass region this analysis excludes cross sections greater than $\sigma_{p}^{SI} = 1.41 \times 10^{-4}$~pb (90\% C.L.). The CRESST collaboration has also reported the observation of an excess of events with a best fit for a dark matter particle with a mass of $\sim 13$~GeV/c$^{2}$ and a cross section of $3 \times 10^{-5}$~pb~\cite{Angloher:2011uu}. Furthermore, this mass range is similar to that required to explain the spectrum of $\gamma$ radiation observed by FERMI from the galactic centre~\cite{Hooper2011412}. 

\begin{figure}[htb]
\begin{center}
\includegraphics[width=.8\textwidth]{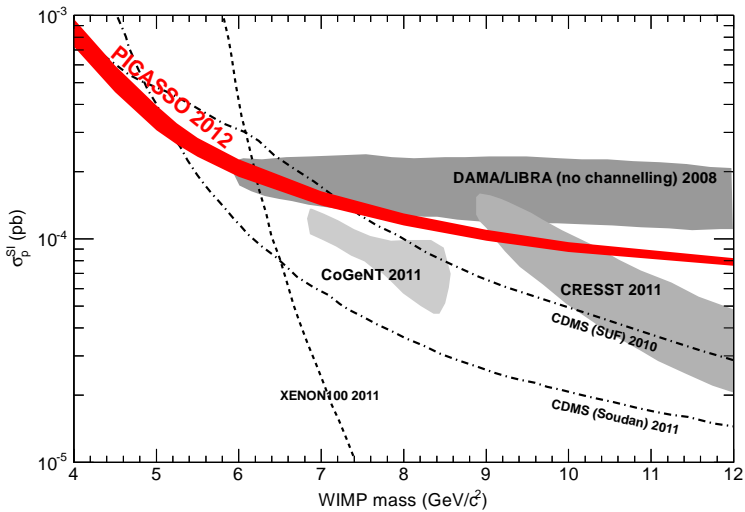}
\end{center}
\caption{PICASSO limits in the spin independent sector (90\% C.L.). Only the region of recent interest in the range of low WIMP masses is shown. The allowed regions of DAMA/LIBRA\cite{springerlink:10.1140/epjc/s10052-010-1303-9}, CoGeNT~\cite{PhysRevLett.106.131301} and CRESST~\cite{Angloher:2011uu} and the exclusion limits by XENON100~\cite{PhysRevLett.107.131302} and CDMS~\cite{PhysRevLett.106.131302} are shown. The broadening of the PICASSO exclusion limit is due to the uncertainty in the energy resolution at low threshold energies.}\label{Fig:7}
\end{figure}

A summary of allowed regions and exclusion limits in the low mass region is shown in Fig.~\ref{Fig:7}. The broadening of the PICASSO exclusion limit is due to the increasing effect of the uncertainty in the energy resolution parameter, $a$, in the low mass region. The interpretation of the  DAMA/LIBRA modulation effect shown in Fig.~\ref{Fig:7} in terms of evidence of interactions of dark matter particles with $^{22}$Na nuclei assumes a quenching factor of $Q_{Na} = 0.3$. It is interesting to note that this allowed region appears to be disfavored by PICASSO using a target nucleus of an atomic weight very close to that of $^{22}$Na.   

%%%%%%%%%%%%%%%%%%%%%
\section{Summary and Perspectives}
\label{sap}
The analysis of 10 detectors in the PICASSO set-up at SNOLAB resulted in exclusion limits on spin dependent interactions of dark matter particles with protons of $\sigma_{p} = 0.032$~pb at 90\% C.L for a WIMP mass of 20~GeV/c$^{2}$.  These limits are more stringent by a factor five than the previous PICASSO 2009 results and with the normal model for WIMP interactions rule out the ion channelling hypothesis invoked to explain the DAMA/LIBRA modulation effect. The use of the light target nucleus $^{19}$F, combined with the low detection threshold of 1.7~keV for recoil nuclei, renders PICASSO particularly sensitive to low mass dark matter particles and gives it also some leverage in the low mass region of the spin independent sector. The present stage of the experiment is approaching the sensitivity to challenge or confirm the claims of seasonal modulations by the DAMA and CoGeNT experiments. 

The main improvements with respect to our previous published results are: a reduction in $\alpha$ background by up to a factor eight due to improvements in detector purification and fabrication; use of a new discrimination variable allowing efficient discrimination of non-particle induced events at low recoil energy thresholds;  and the extension of the analysis from 2 to 10 detectors. 

In the current 32 detector set up eight additional modules have low enough background to be used in the standard analysis described here and will be included in the analysis, once their exposure gives them sufficient statistical weight. Detector modules with higher background will be gradually replaced by cleaner modules depending on progress in detector fabrication and purification. 

The implementation of the event by event $\alpha$ recoil discrimination using the acoustic signal energy discovered by PICASSO and described in~\cite{1367-2630-13-4-043006, 1367-2630-10-10-103017} is proceeding and will allow a substantial increase of sensitivity. In order to match the anticipated sensitivity of the next stage of PICASSO, the experiment has been moved to a new location at SNOLAB, allowing an expansion of water shielding with a substantial improvement in neutron suppression.    

%%%%%%%%%%%%%%%%%%%%%
\section*{Acknowledgements}
\label{ack}
We wish to acknowledge the support of the National Sciences and Engineering Research Council of Canada (NSERC), the Canada Foundation for Innovation (CFI) and the National Science Foundation (NSF 0856273). We also acknowledge support from the Department of Atomic Energy (DAE), Govt.\ of India, under the project CAPP at SINP, Kolkata and the Czech Ministry of Education, Youth and Sports within the project MSM6840770029. We thank J. Behnke and A. Grandison for technical support. We wish to give special thanks to SNOLAB and its staff for their hospitality and for providing help and advice whenever needed.



%% References


\bibliographystyle{model1-num-names}
\bibliography{picasso_2012.bib}






% \end{thebibliography}

\end{document}